\documentclass[a4paper,11pt,dvitex]{article}
\usepackage{pos}

\title{Phase structure of lattice QCD in the heavy quark 
high-density region and the three-state Potts model}
 \ShortTitle{Phase structure of lattice QCD in the heavy
dense region and the three-state Potts model}

\author*[a]{Shinji Ejiri}
\author[b]{Masanari Koiida}
\author[b]{Toshiki Sato}

\affiliation[a]{Department of Physics, Niigata University, Niigata 950-2181, Japan}
\affiliation[b]{Graduate School of Science and Technology, Niigata University, Niigata 950-2181, Japan}

\emailAdd{ejiri@muse.sc.niigata-u.ac.jp}
\abstract{
We discuss the nature of the QCD phase transition in the heavy quark high-density region by considering an effective theory in which Polyakov loops are dynamical variables.
The Polyakov loop is an order parameter of $Z_3$ symmetry, and the fundamental properties of the phase transition are thought to be determined by the $Z_3$ symmetry broken by the phase transition.
By replacing the Polyakov loop with $Z_3$ spin, we find that the effective model becomes a three-dimensional three-state Potts model ($Z_3$ spin model) with a complex external field term.
We investigate the phase structure of the Potts model and discuss QCD in the heavy quark region.
The critical points are determined by finite volume scaling analysis, and in the region where the sign problem is severe, the tensor renormalization group is used to investigate.
As the density varies from $\mu=0$ to $\mu=\infty$, we find that the phase transition is first order in the low-density region, changes to a crossover at the critical point, and then becomes first order again. 
This strongly suggests the existence of a first order phase transition in the high-density heavy quark region of QCD.
}

\FullConference{The 42nd International Symposium on Lattice Field Theory (LATTICE2025)\\
2-8 November 2025\\
Tata Institute of Fundamental Research, Mumbai, India\\}

\begin{document}
\maketitle

\section{Introduction}
\label{sec:intro}

The nature of the finite temperature phase transition of QCD depends on the particle density and the mass of the dynamical quarks. 
It is well known that in the case of infinite quark mass, i.e., quenched QCD, the finite temperature phase transition is first order, and as the mass decreases the phase transition changes to a crossover at a critical mass. 
Although there is much discussion, the standard understanding is as follows:
In the massless limit, the phase transition in two-flavor QCD is second-order, and in the case of three or more flavors, the phase transition is first order.
There is a first-order region near the three-flavor massless limit.

The fundamental properties of a phase transition are thought to be determined solely by the symmetry broken during the phase transition and the dimension of the space.
In the early days of finite temperature lattice QCD, the three-state Potts model, which has the same broken symmetry, was used to discuss how the nature of the phase transition changes when the effects of dynamical quarks are added from quenched QCD \cite{DeGrand:1983fk}.
When the quark mass is sufficiently large, the hopping parameter expansion leads to the effective action of the quark, whose leading term is proportional to the Polyakov loop,
which is the order parameter of the $Z_3$ center symmetry that is broken at the QCD phase transition.
The Polyakov loop corresponds to the $Z_3$ spin, and the quark determinant corresponds to the external magnetic field term in the spin model.

In this study, we apply this argument to investigate the phase structure of high-density lattice QCD in comparison with the spin model.
There is an effective theory that describes the heavy-quark high-density limit of QCD \cite{Blum:1995cb,Aarts:2017vrv}.
The effective theory is a model in which the Polyakov loop is the dynamical variable and the Boltzmann factor of a quark is controlled only by one parameter $C(\mu, m_q)$, a function of the quark mass $m_q$ and the chemical potential $\mu$.
In Sec.~\ref{sec:hdqcd}, we show that by replacing the Polyakov loop with a $Z_3$ spin, the effective model becomes a three-dimensional three-state Potts model ($Z_3$ spin model) with a complex external magnetic field term.
We investigate the phase structure of the Potts model corresponding to QCD in the heavy quark region and how the nature of the phase transition changes when the chemical potential is increased.
In Sec.~\ref{sec:critical}, we determine the critical point where the first-order phase transition ends in the three-state Potts model corresponding to QCD.
In Sec.~\ref{sec:crossover}, we discuss the behavior of physical quantities in the crossover region using the tensor renormalization group method.
Section \ref{sec:summary} provides conclusions.

\section{Heavy dense QCD and three-state Potts model}
\label{sec:hdqcd}

\paragraph{Hopping parameter expansion}

The partition function of lattice QCD with $N_{\rm f}$ flavors of quarks is
${\cal Z} =\int{\cal D} U (\det M)^{N_{\rm f}} e^{-S_{\rm g} (U)},$
where $U_{\mu}(x)$ is the $SU(3)$ link field, $S_g$ is the action of the gauge field, and 
$M$ is the quark kernel on a lattice with the size $N_s^3 \times N_t$.
We adopt the standard Wilson fermion.
To investigate the case of large quark masses, we expand the quark determinant in terms of the hopping parameter $\kappa$:
\begin{eqnarray}
\ln \det M(\kappa) = \ln \det M(0) 
+ N_{\rm site} \sum_{n=1}^{\infty} \sum_{m= -\infty}^{\infty} L_m (N_t, n) e^{m \mu /T} \kappa^n,
\end{eqnarray}
where $\mu$ is the chemical potential, and 
the temperature $T$ is defined as the inverse of the length in the time direction: $T=1/(N_t a)$.
$a$ is the lattice spacing.
$m$ is the winding number due to the periodic boundary condition in the time direction.
The sign of $m$ indicates whether the winding is in the positive or negative direction, and satisfies $L_{-m} = L_m^*$.

These expansion coefficients have been calculated on configurations generated near the phase transition point in Ref.~\cite{Wakabayashi:2021eye,Ejiri:2023tdp} and found that 
$L_m (N_t, n)$ for $m \geq 2$ are much smaller than $L_1 (N_t, n)$.
Moreover, for $m=\pm1$, $L_1 (N_t, n)$ is proportional to the standard linear Polyakov loop $\Omega$, i.e.,
$L_1 (N_t, n) = L_1^0 (N_t, n) c_n \Omega$, where $L_1^0 (N_t, n) c_n$ is the proportional constant. 
Thus, 
\begin{eqnarray}
\sum_{n=N_t}^{\infty} L_1 (N_t, n) e^{\mu/T} \kappa^n 
\approx \Omega e^{\mu/T} \sum_{n=N_t}^{\infty} L_1^0 (N_t, n) c_n \kappa^n
\equiv 6 \times 2^{N_t} N_t^{-1} \Omega e^{\mu/T} \kappa_{\rm eff}^{N_t} .
\end{eqnarray}
Here, the right-hand side is the leading term $L_1(N_t, N_t) e^{\mu/T} \kappa^{N_t}$ 
with $\kappa$ replaced by $\kappa_{\rm eff}$.
This means that by shifting $\kappa$ to $\kappa_{\rm eff}$ in the relatively wide $\kappa$ region where the hopping parameter expansion is valid, the effects of higher-order terms $L_m(N_t, n)$, including spatial links, can be incorporated into an expansion of $\det M$, which ignores terms including spatial links.
In lattice QCD, the theory that approximates it by increasing the quark mass and ignoring the spatial link term is called the heavy dense effective theory of QCD \cite{Blum:1995cb,Aarts:2017vrv}.
Here we think it is worthwhile to revisit the heavy dense effective theory and discuss QCD at high densities by shifting $\kappa$ to $\kappa_{\rm eff}$.

\paragraph{Heavy quark high-density effective theory}

When the quark mass is heavy and the chemical potential is large, i.e. $\kappa \rightarrow 0, e^{\mu a} \rightarrow \infty$, the quark kernel can be simplified to
\begin{eqnarray}
M_{x,y}=\delta_{x,y} - \kappa(1-\gamma_4)U_4 (x) e^{\mu a} \delta_{y,x+\hat{4}}.
\end{eqnarray}
Since there is no link field in the spatial direction, the quark determinant is expressed as a product of the determinants at each spatial point $\vec{x}$.
Then, the quark determinant can be rewritten by
\begin{eqnarray}
\mathrm{det}M = \prod_{\vec{x}} \left\{1+3C \Omega^{\rm (loc)} (\vec{x}) + 3C^2 \left( \Omega^{\rm (loc)} (\vec{x}) \right)^* + C^3 \right\}^2.
\end{eqnarray}
Here, we introduce the parameter $C=(2\kappa)^{N_t} e^{\mu/T},$ and 
$\Omega^{\rm (loc)} (\vec{x})$ is the local Polyakov loop at each point $\vec{x}$ defined as
$\Omega^{\rm (loc)} (\vec{x}) = \frac{1}{3} \mathrm{tr}\prod_{i=0}^{N_t-1} U_4 (\vec{x}+i\hat{4}).$

This effective theory is symmetric under transformations that transform $C$ into $C^{-1}$ and simultaneously exchange $\Omega^{\rm (loc)}(\vec{x})$ and $\Omega^{\rm (loc)*}(\vec{x})$ \cite{Blum:1995cb}, where we ignore overall constants that do not affect the calculation of the path integral.
The exchange of complex conjugates $\Omega^{\rm (loc)} (\vec{x})$ is the exchange of particles and antiparticles, and the transformation from $C$ to $C^{-1}$ is a swap of low density and high density.
In fact, when $C$ is infinite, the $\Omega^{\rm (loc)}$ and $\Omega^{\rm (loc)*}$ terms become ineffective and the quark determinant becomes a constant, which is the same as in quenched QCD with $C=0$.
If the critical point is found at $C_c$, due to this symmetry, there will be two critical chemical potentials $\mu_c$, with $\mu_c/T = \pm \ln C_c - N_t \ln(2 \kappa)$.
Decreasing the quark mass (increasing $\kappa$) decreases both the large and small critical chemical potentials.
The critical chemical potential in the high-density region may be related to the critical point in the light-quark, low-density region. 

\paragraph{Effective three-dimensional spin model}

Since the properties of the phase transition are considered to be the same if the symmetry broken in the phase transition is the same, we consider the three-dimensional three-state Potts model. 
The effective model of heavy dense QCD and the Potts model both break $Z_3$ symmetry at the phase transition.
Similar to the early days of lattice QCD when the effect of dynamical quarks on the finite temperature phase transition was discussed in comparison with the Potts model \cite{DeGrand:1983fk}, we replace the Polyakov loop with a $Z_3$ spin variable.
The integral measure and gauge field action are replaced by the spin variable, $s(\vec{x})$: \\
%\begin{eqnarray}
\hspace{7mm}
$3 \Omega^{\rm (loc)}  (\vec{x}) \to s  (\vec{x}), \hspace{5mm}
\int {\cal D} U \ \rightarrow \ \sum_{s(\vec{x})} , \hspace{5mm}
S_{\rm g} (U) \ \rightarrow \ -\beta\sum_{\vec{x}} \sum_{i=1}^3
\mathrm{Re} \left[ s (\vec{x}) s^* (\vec{x}+\hat{i}) \right].$ \\
%\end{eqnarray}
Here, $s (\vec{x})$ can take the following three states:
$s_1=1$, $s_2=e^{2\pi i/3}$, and $s_3=e^{-2\pi i/3}$.
In the absence of an external magnetic field, this model has $Z_3$ symmetry.
We show that the equivalent of the quark determinant is the complex external magnetic field term.
Replacing $3 \Omega^{\rm (loc)}  (\vec{x})$ with $s  (\vec{x})$, \\
\hspace{20mm} 
%\begin{eqnarray}
$(\det M)^{N_{\rm f}} \rightarrow \prod_{\vec{x}}(1+C s (\vec{x}) + C^2 s^* (\vec{x}) + C^3 )^{2N_{\rm f}}
\equiv \prod_{\vec{x}} F (\vec{x}) .$ \\
%\end{eqnarray}
If $s (\vec{x})=s_1$,
$F (\vec{x}) = e^{A_1};$
if $s (\vec{x})=s_2$,
$F (\vec{x}) = e^{A_2 +i\theta};$
if $s (\vec{x})=s_3$, 
$F (\vec{x}) = e^{A_2 - i\theta}.$
Assuming the number of spins taking state $s_i$ is $N_i$ and the total number of spins is $N_{\rm site}=N_1+N_2+N_3$, from the above equations, we get
$\prod_{\vec{x}} F (\vec{x}) = \exp [N_1A_1+N_2(A_2 + i\theta)+N_3(A_2 - i\theta)] .$
Furthermore, from the relationship between $N_i$ and $s_i$:
$\sum_{\vec{x}} \mathrm{Re} \left[ s (\vec{x}) \right] = N_1 - \frac{1}{2}(N_2 + N_3) 
= \frac{3}{2} N_1 - \frac{1}{2} N_{\rm site}$ and
$\sum_{\vec{x}} \mathrm{Im} \left[ s (\vec{x}) \right] = \frac{\sqrt{3}}{2} (N_2-N_3)$,
${\cal Z}$ is given by
\begin{eqnarray}
{\cal Z} = \sum_{s(\vec{x})} \exp \left[
\beta\sum_{\vec{x}} \sum_{i=1}^3 \mathrm{Re} \left[ s (\vec{x}) s^* (\vec{x}+\hat{i}) \right]
+h\sum_{\vec{x}} \mathrm{Re} [s (\vec{x})]
+iq\sum_{\vec{x}} \mathrm{Im} [s (\vec{x})]
\right] ,
\label{eq:Zpott}
\end{eqnarray}
excluding the overall constant.
This is the partition function of the three-dimensional three-state Potts model, but with terms for real and imaginary external fields.
When corresponding to heavy dense QCD, the parameters $h$ and $q$ are real numbers and is functions of $C$ as follows:
\begin{eqnarray}
h &=& \frac{4}{3}N_{\rm f} \mathrm{ln}(1+C+C^2+C^3)
- \frac{2}{3}N_{\rm f} \mathrm{ln} \left[ \left(1-\frac{1}{2}C -\frac{1}{2}C^2 +C^3\right)^2 + \frac{3}{4}(C-C^2)^2 \right], 
\label{eq:pottsh}
\\
q &=& \frac{4}{\sqrt{3}}N_{\rm f} \ \mathrm{arctan}\left[ 
\frac{\frac{\sqrt{3}}{2}(C-C^2)}{1-\frac{C}{2}-\frac{C^2}{2}+C^3}
\right].
\label{eq:pottsq}
\end{eqnarray}
In Figure~{\ref{fig:hq2}, the red curve shows the trajectory of $(h, q)$ in parameter space for $N_{\rm f}=2$ as $C$ varies from zero to infinity.
Details of this effective model are given in Ref.~\cite{Ejiri:2026ijj}.

\begin{figure}[tb]
\begin{minipage}{0.47\hsize}
\begin{center}
\vspace{-2mm}
\includegraphics[width=7.0cm]{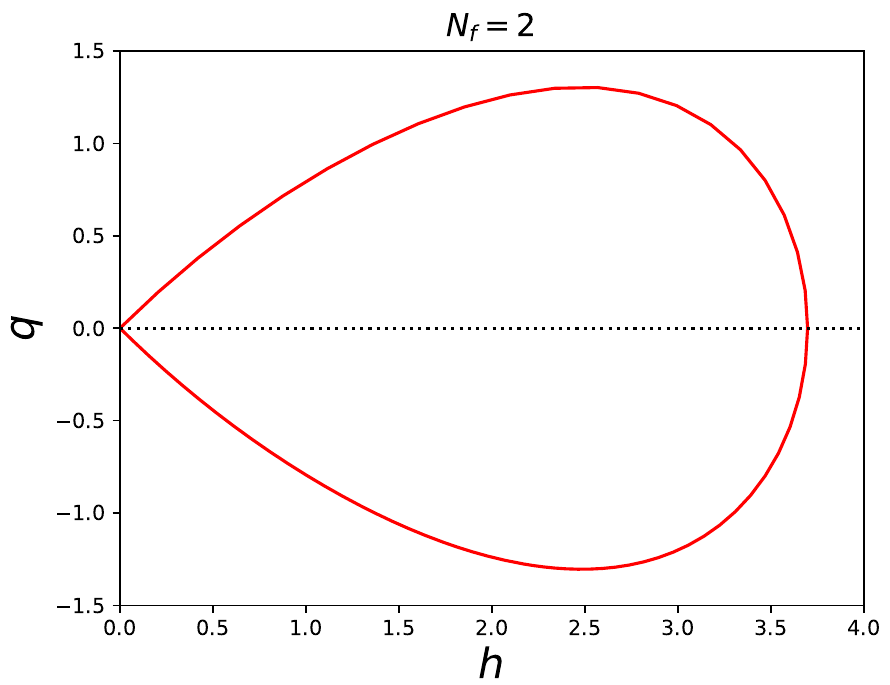}
\vspace{-6mm}
\end{center}
\caption{The corresponding parameters $(h,q)$ of the spin model when changing $C$ in heavy dense QCD for $N_{\rm f}=2$ \cite{Ejiri:2026ijj}.}
\label{fig:hq2}
\end{minipage}
\hspace{2mm}
\begin{minipage}{0.47\hsize}
\begin{center}
\vspace{-3mm}
\includegraphics[width=7.0cm]{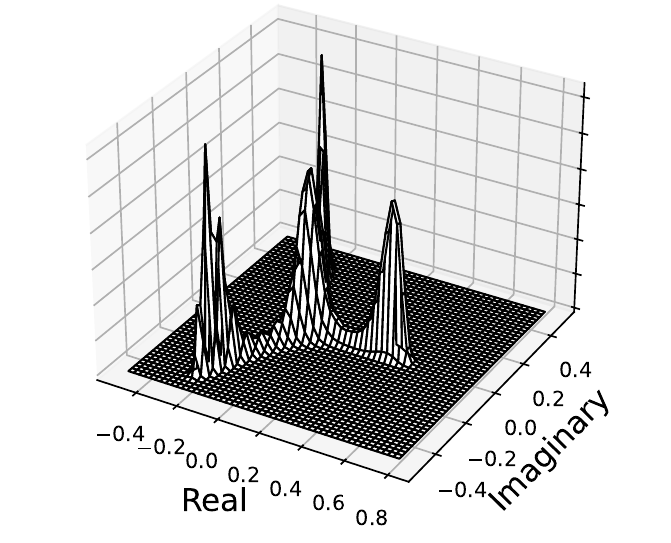}
\vspace{-7mm}
\end{center}
\caption{Probability distribution of magnetization in the complex plane for the three-state Potts model at $(\beta, h, q)=(0.36697, 0.0, 0.0)$ on a $40^3$ lattice.}
\label{fig:shist}
\end{minipage}
\end{figure}

\paragraph{Histogram of magnetization}

\begin{figure}[tb]
\begin{center}
\vspace{0mm}
\includegraphics[width=5.6cm]{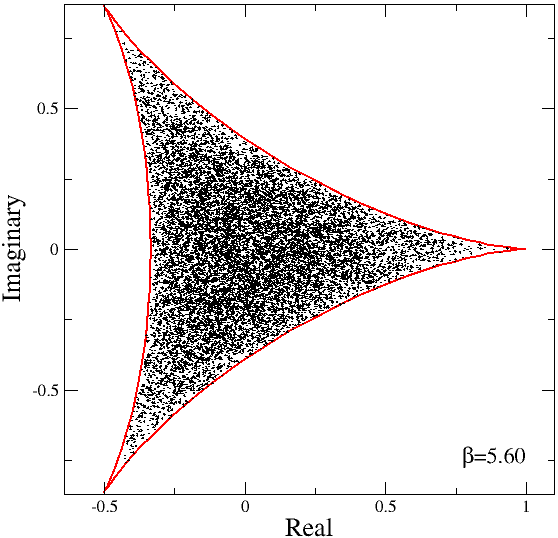}
\hspace{4mm}
\includegraphics[width=5.6cm]{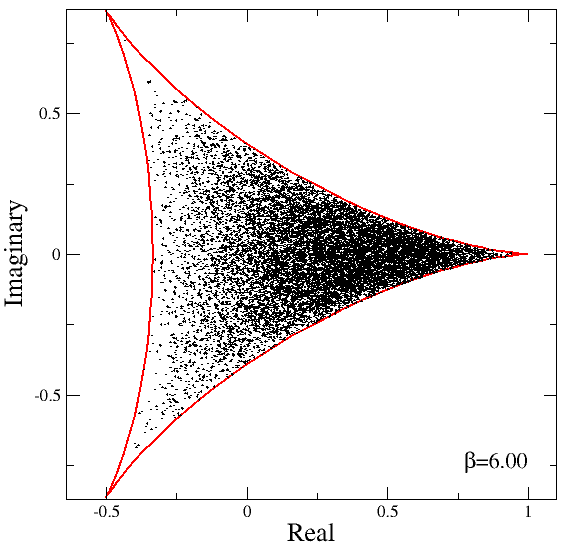}
\vspace{-4mm}
\end{center}
\caption{Distribution of the local Polyakov loop at each point in one configuration of quenched QCD.
The left panel shows the distribution for the symmetric phase $(\beta =5.60)$, and the right panel shows the distribution for the broken phase $\beta = 6.00)$ \cite{Ejiri:2022jai}.}
\label{fig:pl3hist}
\end{figure}

\begin{figure}[tb]
\begin{center}
\vspace{-2mm}
\includegraphics[width=5.6cm]{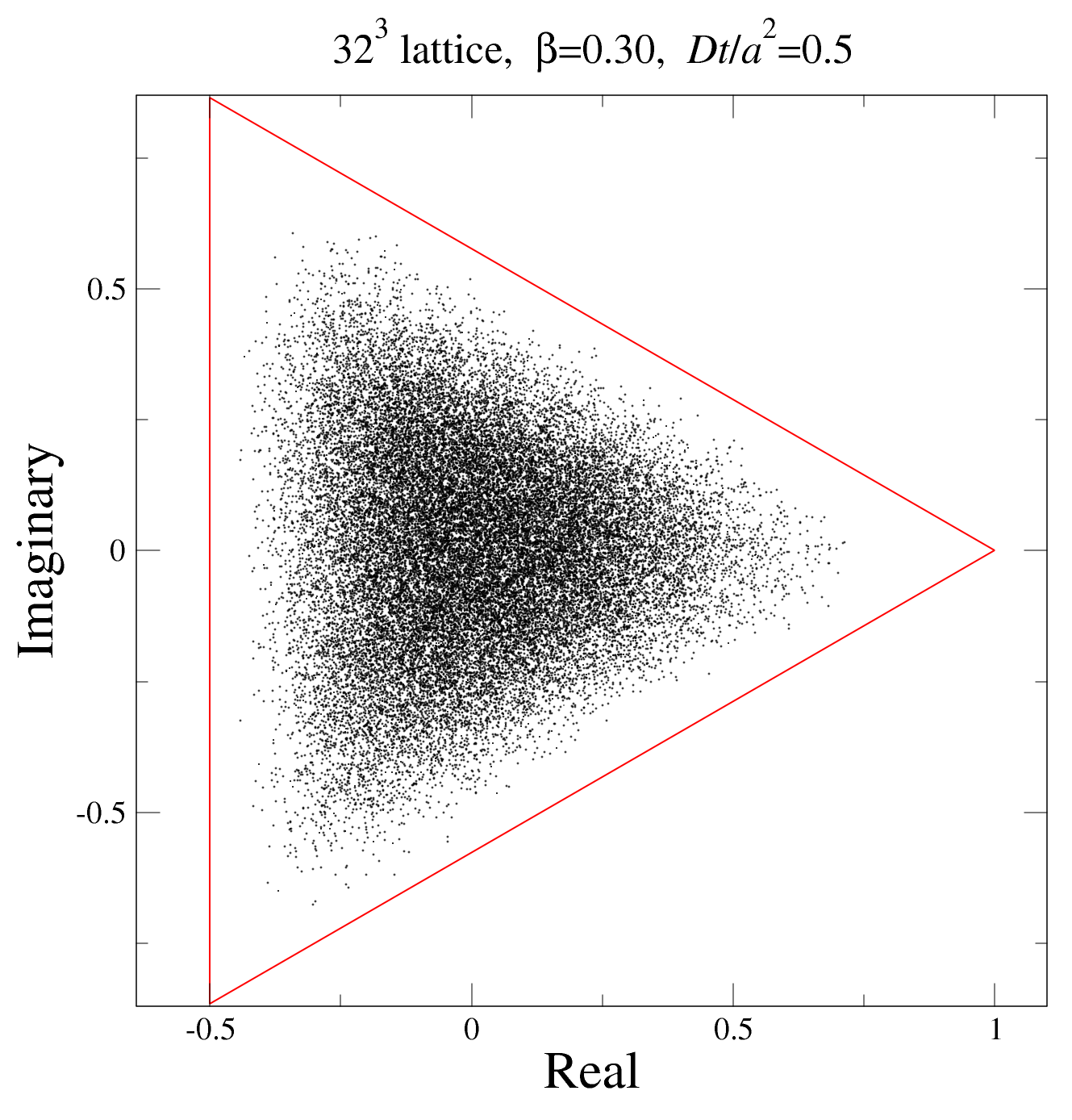}
\hspace{4mm}
\includegraphics[width=5.6cm]{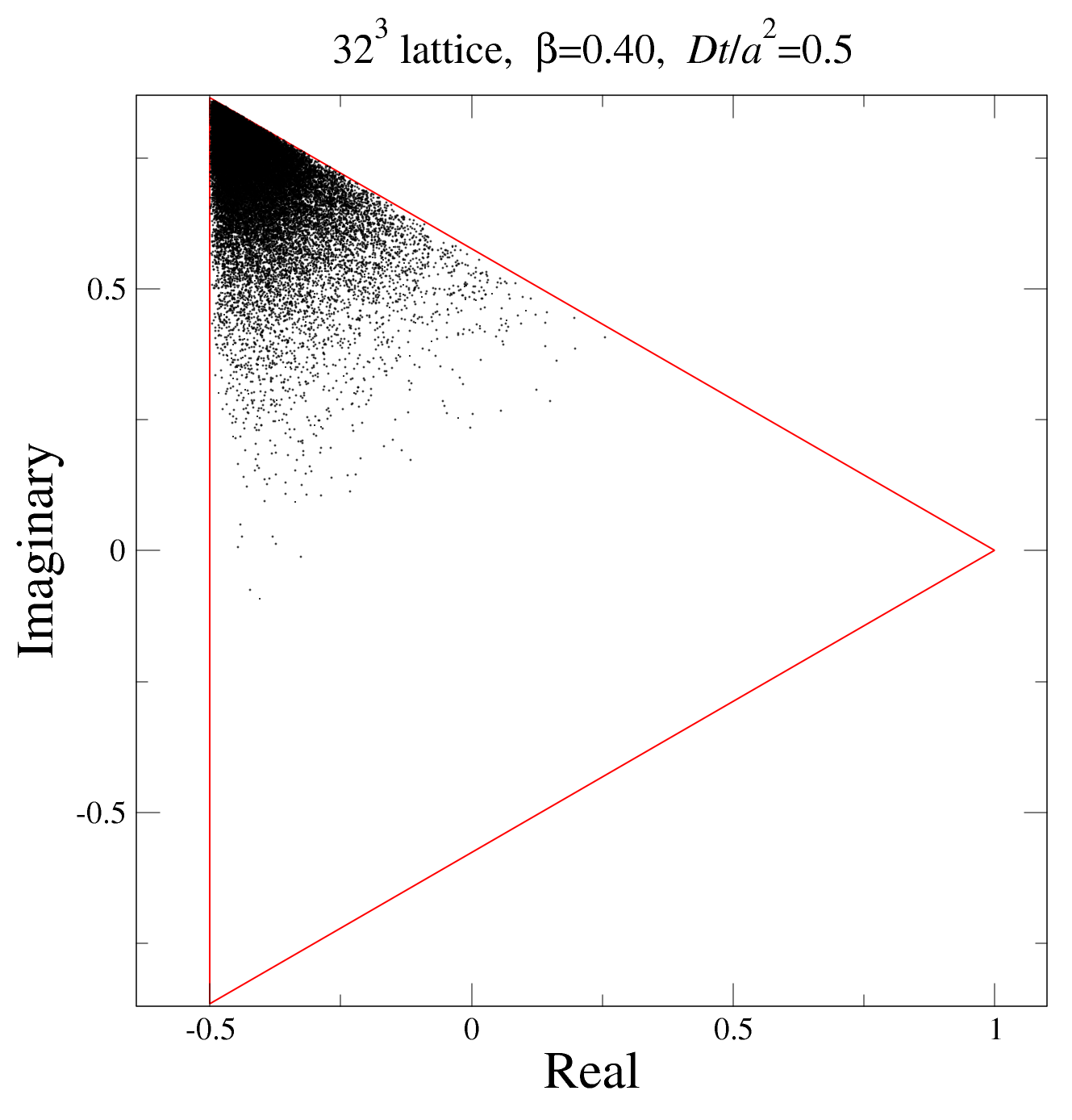}
\vspace{-4mm}
\end{center}
\caption{Distribution of spins $\tilde{s}(\vec{x}, t)$ at each point in one configuration $(\beta=0.30)$ of three-state Potts model, after coarse-graining with the diffusion equation with $Dt/a^2 =0.5$. 
The left panel is $\beta=0.30$ (symmetric phase), and the right panel is $\beta=0.40$ (broken phase) \cite{Ejiri:2026ijj}.}
\label{fig:potdifhist}
\end{figure}

Here we discuss the analogy between the Polyakov loop in QCD and $Z_3$ spin.
The probability distributions of their spatial averages in the complex plane are very similar.
Figure \ref{fig:shist} shows the probability distribution of the spatial average of spin values, i.e., the magnetization, calculated by a simulation of the three-state Potts model at $(\beta, h, q)=(0.36697, 0.0, 0.0)$ on a $40^3$ lattice, which is a typical probability distribution of a first-order phase transition.
This histogram in the complex plane are very similar to the histograms of the Polyakov loop in the heavy-quark limit of QCD \cite{Saito:2013vja}. 

Since $\Omega^{\rm (loc)} (\vec{x})$ is the trace of an $SU(3)$ matrix, its values are distributed within the distorted triangle in the complex plane  enclosed by the red line in Fig.~\ref{fig:pl3hist} \cite{Ejiri:2022jai}.
The horizontal and vertical axes is the real and imaginary parts of $\Omega^{\rm (loc)}$.
The left panel shows the result for the confinement phase $(\beta =5.60)$ and the right panel shows the result for the deconfinement phase $(\beta =6.00)$ calculated by quenched QCD simulations $(h=q=0)$ on a $24^3 \times 4$ lattice.
In the confinement phase, the distribution is $Z_3$ symmetric, whereas in the deconfinement phase, as shown in the right figure, the distribution is skewed towards the right.
On the other hand, $Z_3$ spin is a variable that takes on the values 
$s (\vec{x}) = 1$, $e^{2\pi i/3}$, and $e^{-2\pi i/3}$.
Therefore, symmetry breaking occurs when the number of spins that take on these three values becomes asymmetric.
Now, if we coarse-grain the spin using the diffusion equation:
$ \frac{\partial}{\partial t} \tilde{s} (\vec{x}, t) = D \vec{\nabla}^2 \tilde{s} (\vec{x}, t), $
we get distributions shown in Fig.~\ref{fig:potdifhist}.
$D$ represents the diffusion coefficient.
The left figure shows the result for the symmetric phase, and the right figure shows the result for the broken phase computed on a $32^3$ lattice.
These results are very similar to the distribution of the Polyakov loop at each point $\Omega^{\rm (loc)} (\vec{x})$.
Note that the spatial average of the spin values remains unchanged even after the coarse-graining.
In other words, the external magnetic field terms in the Hamiltonian do not change under the coarse-graining.

\section{Critical points where first-order phase transitions end}
\label{sec:critical}

We study the phase structure of heavy quark high-density QCD by investigating the three-dimensional three-state Potts model given by Eq.~(\ref{eq:Zpott}) \cite{Ejiri:2026ijj}.
We fix the hopping parameter at a small value and then gradually increase the chemical potential from a small value to infinity.
The $(h, q)$ parameters are shown in Fig.~\ref{fig:hq2} for $N_{\rm f} =2$.
When $C=0$, it is quenched QCD, so the phase transition is first-order.
In the standard Potts model with $q=0$, the first-order transition is known to change into a crossover at $h = 0.000517(7)$ \cite{Karsch:2000xv}.
Furthermore, the critical point belongs to the same universality class as the three-dimensional Ising model.
Near $C=0$, $h$ and $q$ are approximately the same, so we expect that at the critical point, $q$ will be small and the sign problem will be mild.
Therefore, we apply the reweighting method for $q$ to perform scaling analysis of the magnetization.

We compute the magnetization:
$ m= (1/N_{\rm site}) \sum_{\vec{x}} \mathrm{Re} [s (\vec{x})] , $
by Monte Carlo simulation of the three-dimensional three-state Potts model while varying the spatial volume $L^3$.
The standard Metropolis algorithm is used.
For each parameter, the number of independent configurations is set to 200000.
We first determine the $\beta$ at which the phase transition occurs, denoted as $\beta_c$, for various $h$, and then calculate the Binder cumulant 
$B_4= \langle (m -\langle m \rangle )^4 \rangle / \langle (m -\langle m \rangle )^2 \rangle^2$ 
at $\beta_c$ on lattices with side lengths $L$ from 40 to 90.
In Fig.~\ref{fig:b4scale}, we plot $B_4$ at $\beta_c$ calculated for various $h$ and $L$.
The coefficient $q$ of the imaginary part of the external field term, which is incorporated by reweighting, satisfies Eqs.~(\ref{eq:pottsh}) and (\ref{eq:pottsq}) for each $h$ with $N_{\rm f}=2$.
It can be seen that $B_4$ as a function of $h$, calculated for all $L$, intersects at a certain $h$.
The point $h$ at which the $L$-dependence disappears is the critical point $h_c$.
In the vicinity of the critical point, $B_4$ satisfies the following equation:
\begin{equation}
B_4 (t,0,L^{-1}) = B_4 (0,0,L^{-1}) + A t L^{1/\nu} +O(t^2).
\end{equation}
$A$ is a proportionality constant.
Assuming that $t=h-h_c$, we fit the data to the equation, 
where the fit parameters are $B_{4c}$, $h_c$, $\nu$, and $A$.
The dashed lines show the fit functions obtained by fitting the data for each $L$.
$B_{4c}$ is the Binder cumulant at the critical point, a quantity that is uniquely determined for each universality class.
We obtain $B_{4c}=1.601(6)$. Our result is consistent with the result of the three-dimensional Ising model, $B_{4c}=1.601$.
The result for the critical point $h_c=0.000479(3)$ of the model corresponding to heavy dense QCD are smaller than the standard Potts model with $q=0$.
Converting to the parameter $C$ of heavy dense QCD, $C$ at the critical point is 
$C_c = 1.195(7) \times 10^{-4}$. For details, see Ref.~\cite{Ejiri:2026ijj}.

Furthermore, we calculate the magnetic susceptibility 
$\chi_m = N_{\rm site} \langle (m - \langle m \rangle )^2 \rangle$ 
varying the spatial volume $L^3$.
A reweighting method is used to account for the effect of finite $q$.
Near the critical point, the magnetic susceptibility is expected to follow the scaling relation:
\begin{equation}
  \chi_m(t,0,L^{-1}) L^{-2+\eta} =  \tilde{f}_2(0,0) + B t L^{1/\nu} +{\cal O}(t^2), 
  \label{eq:susp2fit}
\end{equation}
where $B$ is a proportionality constant.
Assuming $t=h-h_c$, the scaling plot, $(h-h_c) L^{1/\nu}$ vs. $\chi_m L^{-2+\eta}$, is shown in Fig.~\ref{fig:susscale}.
In this plot, the critical exponents $\nu$ and $\eta$ are those of the Ising model, and the critical point $h_c$ is the result obtained from the fit of $B_4$.
From Fig.~\ref{fig:susscale}, we can see that the data are aligned in a straight line near the critical point.
The dashed line is a linear fit to the data near the critical point.
This scaling plot shows that the phase transition of this model belongs to the same universality class as the three-dimensional Ising model.
A similar analysis can be extended to the case where the chemical potential is complex.
For real $\mu$, the results are not very different from the standard Potts model, but when $\mu$ is a complex value, a novel singularity is found \cite{Ejiri:2026ijj}.

\begin{figure}[tb]
\begin{minipage}{0.47\hsize}
\begin{center}
\vspace{-2mm}
\includegraphics[width=7.0cm]{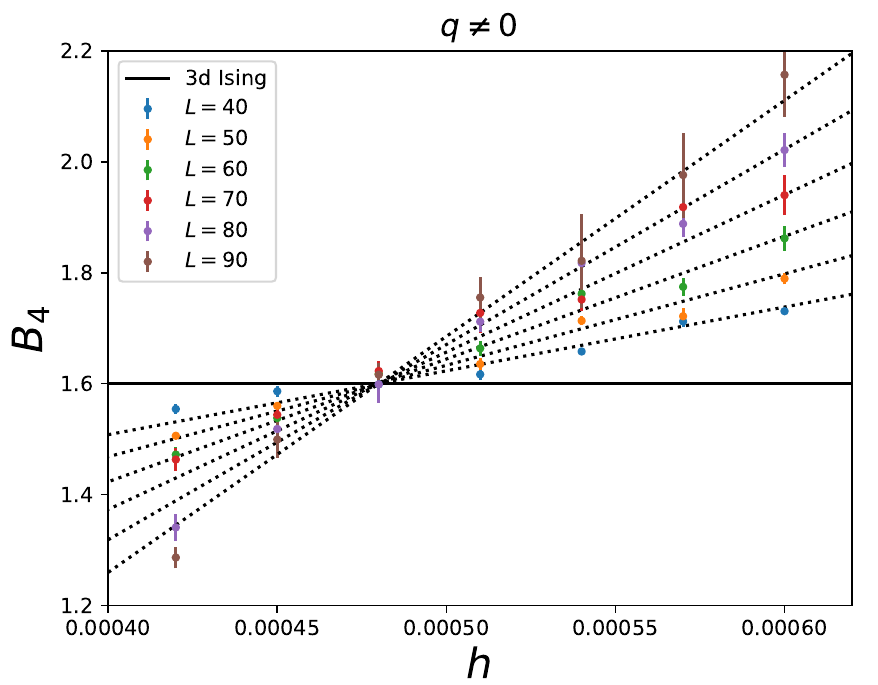}
\vspace{-6mm}
\end{center}
\caption{Binder cumulant at the $\beta_c$ as a function of $h$ on lattices with $L$ from 40 to 90. }
\label{fig:b4scale}
\end{minipage}
\hspace{2mm}
\begin{minipage}{0.47\hsize}
\begin{center}
\vspace{-2mm}
\includegraphics[width=7.0cm]{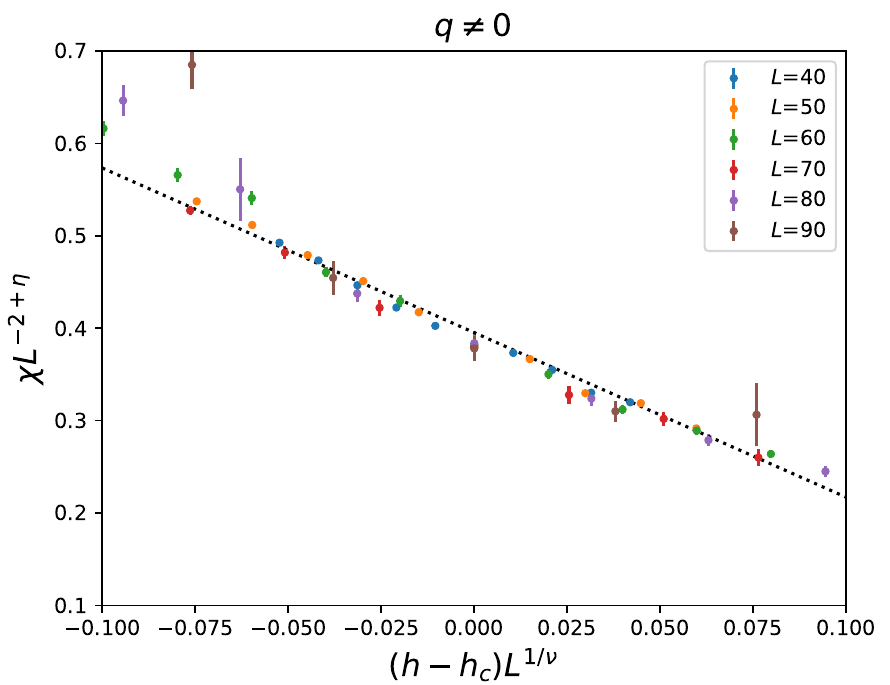}
\vspace{-6mm}
\end{center}
\caption{Scaling plot of the peak height of the magnetic susceptibility.}
\label{fig:susscale}
\end{minipage}
\end{figure}

\begin{figure}[tb]
\begin{center}
\vspace{-2mm}
\includegraphics[width=7.3cm]{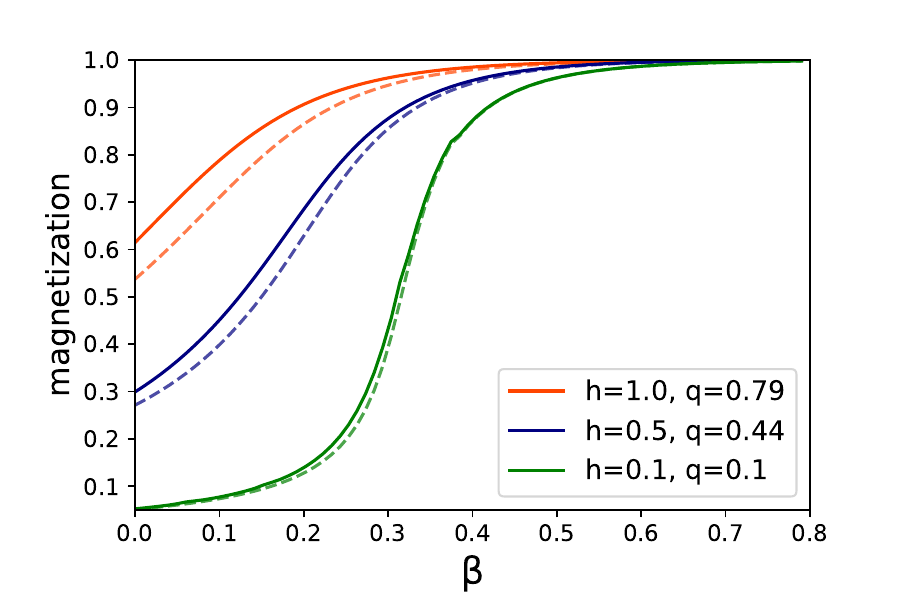}
\hspace{0mm}
\includegraphics[width=7.3cm]{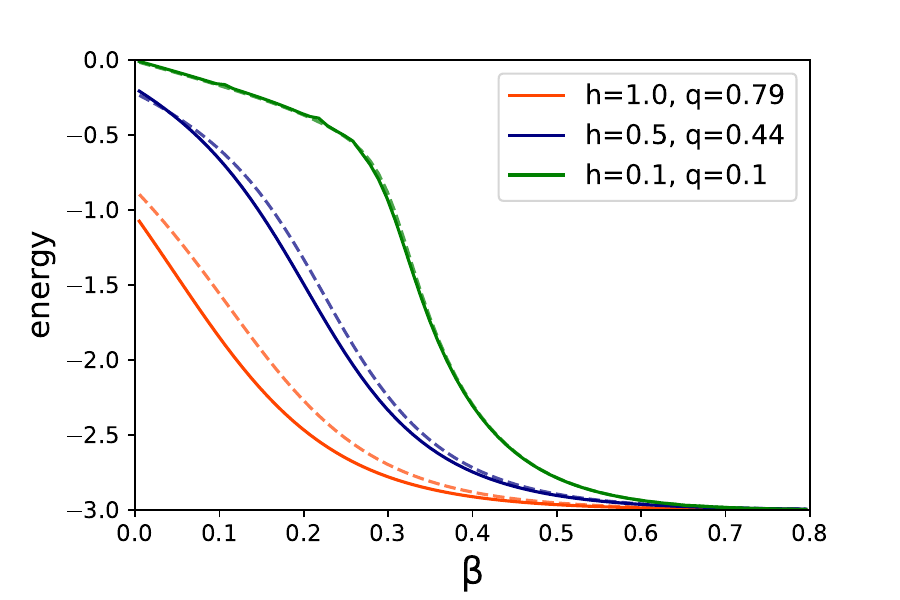}
\vspace{-6mm}
\end{center}
\caption{Magnetization (left) and energy density (right) as functions of $\beta$ computed by HOTRG.}
\label{fig:energb}
\end{figure}

\section{Magnetization and energy at high density by tensor renormalization group}
\label{sec:crossover}

Next, we investigate the behavior of physical quantities in the region where $h$ and $q$ are large beyond the critical point.
When the imaginary part $q$ of the external magnetic field term becomes large, the sign problem becomes serious.
Because this effective theory is a three-dimensional model, the tensor renormalization group method can be applied.
In the tensor renormalization group, calculations become more difficult as the spatial dimension increases, and therefore the use of the tensor renormalization group is one of the motivations for our study to constructing three-dimensional effective theories.

We apply the higher-order tensor renormalization group (HOTRG) method \cite{Xie:2012mjn,Morita:2018tpw}.
Instead of the spin variable $s(\vec{x})= \{1, e^{2 \pi/3}, e^{2 \pi/3} \}$, we set the spin values to $s_i= \{ 0, 1, 2 \}$, 
where $i$ represents the position of the spin.
The local tensor is introduced as follows:
\begin{eqnarray}
T_{x_i x'_i y_i y'_i z_i z'_i} \hspace{-3mm} &=& \hspace{-3mm} \sum_{s_i} 
\left( Q_{s_i x_i} Q^*_{s_i x'_i} Q_{s_i y_i} Q^*_{s_i y'_i} Q_{s_i z_i} Q^*_{s_i z'_i} 
\times \exp \left[ \frac{3}{2} h \delta_{s_i, 0} + \frac{\sqrt{3}}{2} iq (\delta_{s_i , 1} - \delta_{s_i , 2}) \right]
\right), \hspace{4mm} \\
Q_{s_i x_i} \hspace{-3mm}  &=& \hspace{-3mm} 
e^{\frac{2 \pi i}{3} x_i s_i} \sqrt{ \frac{e^{3 \beta /2} -1+3 \delta_{x_i,0}}{3}} .
\hspace{8mm}
e^{\frac{3}{2} \beta \delta_{s_i s_j} } = \sum_{t=0}^2 Q_{s_i t} Q^*_{s_j t}.
\end{eqnarray}
Assuming $x, y, z, x', y',$ and $z'$ also take the values $\{ 0, 1, 2 \}$, the partition function: 
\begin{equation}
{\cal Z} = {\rm Tr} \prod_i T_{x_i x'_i y_i y'_i z_i z'_i}, 
\end{equation}
is equal to Eq.~(\ref{eq:Zpott}) except for the constant factor.
We perform a block spin transformation, reduce the number of tensors, and compute the partition function.
Let $(n)$ be the number of blockings,
we define
$M^{(n)}_{x x' y y' z z'} = \sum_k T^{(n)}_{x_1 x'_1 y_1 y'_1 z k} T^{(n)}_{x_2 x'_2 y_2 y'_2 k z'}$.
We combine indices in the $(x, y)$ direction such that $x= x_1\otimes x_2, y= y_1\otimes y_2$.
Then, the number of elements in $x, y$ increases as the square of those numbers.
Therefore, to reduce the number of elements, we separate the indices of $M$ into $x$ and non-$x$ and perform singular value decomposition so that $M M^{\dagger} = U \Lambda U^{\dagger}$ to obtain a unitary matrix $U$ for $x$. 
Here, $\Lambda$ is a diagonal matrix with the eigenvalues arranged in descending order on the diagonal elements.
Since $U^{\dagger} M M^{\dagger} U = \Lambda$, the approximation of the diagonal elements of $\Lambda$ such that the largest $D_{\rm cut}$ elements are nonzero and the rest are zero means that the elements of $x$ in $U_{i x}$ from $1$ to $D_{\rm cut}$ are left and the rest are truncated.
Moreover, if we obtain a similar unitary matrix $V$ for $y$ by singular value decomposition, we obtain a new tensor that is contracted in the $z$ direction:
\begin{eqnarray}
T^{(n+1)}_{x x' y y' z z'} = \sum_{i j k l} U^{(n+1)}_{i x} V^{(n+1)}_{k y} M^{(n)}_{i j k l z z'} U^{(n+1)}_{j x'} V^{(n+1)}_{l y'} 
\end{eqnarray}
We specify $D_{\rm cut}$ and truncate the elements of indices $x, y$ of $T^{(n+1)}$ at $D_{\rm cut}$-th.
By repeating this block spin transformation $3n$ times in each direction, we obtain the partition function on a $(2^n)^3$ lattice.
Since the physical quantities are obtained as derivatives of $\ln {\cal Z}$, we calculate ${\cal Z}$ by slightly shifting $h$ and $\beta$, and then calculate the magnetization: $m = (1/N_{\rm site}) \partial \ln {\cal Z}/ \partial h$, and energy density: $\varepsilon = -(1/N_{\rm site}) \sum_{\vec{x}} \sum_{j=1}^3 \mathrm{Re} \left[ s (\vec{x}) s^* (\vec{x}+\hat{j}) \right] = -(1/N_{\rm site}) \partial \ln {\cal Z}/ \partial \beta$.

We calculate the expectation values of the energy and magnetization in the presence of a complex external field using the HOTRG.
The expectation value of the magnetization, which is the order parameter, is shown in the left panel of Fig.~\ref{fig:energb}. 
The lattice size is $N_{\rm site}=1024^3$. The truncation is $D_{\rm cut}=8$.
The results when $h$ and $q$ are given by Eqs.~(\ref{eq:pottsh}) and (\ref{eq:pottsq}) with $N_{\rm f}=2$ are shown by solid lines, and for comparison, the cases where $q=0$ are shown by dashed lines.
The horizontal axis is $\beta$ and $h$ is set to $0.1$(green), $0.5$(blue), and $1.0$(red).
In the region away from the critical point calculated here, the $D_{\rm cut}$ dependence is negligible.
First, we can see that the effect of $q$ (imaginary part of the external field term) is small.
And when $h$ becomes larger beyond the critical point, the change becomes milder.
Figure~\ref{fig:energb} shows that  the phase transition point $\beta_c$ decreases as $h$ increases, and eventually the phase transition itself disappears.
Similarly, we plot the expectation value of the energy density $\varepsilon$ in the right panel of Fig.~\ref{fig:energb}.
As with the magnetization, the change in the energy density becomes slower as $h$ increases.

When the hopping parameter $\kappa$ in QCD is small, $C$ is a monotonically increasing function of $\mu$,
i.e., $C=(2\kappa)^{N_t} e^{\mu/T}$.
When $C$ is increased from $C=0$, the phase transition is first order up to the critical point, after which it becomes a crossover.
Since $h$ increases up to $C=1$, as $C$ increases further the change in the order parameter becomes milder and eventually the phase transition disappears.
After $C=1$, $h$ becomes smaller and the phase transition becomes stronger as $C$ approaches the second critical point.
Once the critical point is passed, the phase transition is first order up to infinity.
In Sec.~\ref{sec:critical}, we found the critical points are $C_c = 1.195(7) \times 10^{-4}$ and 
$C_c=[1.195(7)]^{-1} \times 10^{4}$.
Since the critical chemical potential is given by $\mu_c/T = \ln C_c - N_t \ln(2 \kappa)$, increasing $\kappa$ decreases $\mu_c$.
When the quark mass is reduced (when $\kappa$ is increased), the smaller $\mu_c$ quickly becomes zero, but it is interesting to see how the larger $\mu_c$ changes as the mass decreases.
The larger $\mu_c$ is related to the filling of space with quarks, since 
when $C$ is infinite, the number of quarks reaches a maximum and fills space.
Thus, the critical point in the large $C$ region may not be expected to be directly connected to the 
experimentally interesting critical point in finite-density QCD for physical quark masses.

If $C$ is small, the quark determinant in heavy dense QCD is
$\det M \approx \exp (6C N_s^3 \Omega)$, 
where $\Omega$ is the Polyakov loop.
An approximation that ignores the complex phase part of $\det M$ allows Monte Carlo simulations.
In Ref.~\cite{Ejiri:2025fsf}, we performed Monte Carlo simulations for the lattice action:
$S=S_g + {\rm Re} \ln \det M = 6\beta N_s^3 N_t P + 6C N_s^3 {\rm Re} \Omega$, 
where $P$ is the plaquette.
We found that as $C$ increases, the transition point $\beta_c$ decreases, and the change in the expectation value of the plaquette becomes steeper.
Although we must add the effect of the complex phase of $\det M$, this behavior suggests the existence of a new singularity.
On the other hand, in the three-state Potts model, no new singularities appear with increasing $C$ or $h$.
In the process of simplifying from the heavy dense QCD to the three-state Potts model as the effective theory, the property of the rapid change of the plaquette seems to have been lost.

\section{Conclusions and outlook}
\label{sec:summary}

We discussed the phase structure of dense QCD with heavy dynamical quarks through the effective theory of the three-state Potts model.
The Potts model has a complex-valued external field, and we investigated the phase transition with parameters corresponding to heavy dense QCD.
At low densities, the transition is first order, but as the density increases, the transition changes to a crossover at a critical point, weakening the transition.
As the density increases further, another critical point appears, and above that point, the transition becomes first order.
The critical points belongs to the three-dimensional Ising universality class.

In the heavy quark high-density limit, the quark determinant becomes a constant, the same as in quenched QCD.
Therefore, the high-density limit of QCD becomes a first-order phase transition.
In the high-density limit, quarks fill up space.
The second critical point at high densities is thought to be related to this filling of space with quarks.
If this is the case, it is likely unrelated to the first-order phase transition at finite density that is of interest in experiments.
We were interested in whether singularities would appear in the region where $h$ is large in the crossover region.
However, we found that singularities do not appear when the external field is large in the three-state Potts model with a complex external field examined in this study.
The Potts model, which corresponds to high-density QCD, is a three-dimensional theory, so it is easy to analyze, and in particular, it can be analyzed using the tensor renormalization group, which does not have the sign problem.
\paragraph{Acknowledgments}
This work is supported by JSPS KAKENHI (Grant Nos. JP21K03550, JP25K07299).

\end{document}